\documentclass[aps,twocolumn,epsfig,graphics,floatfix,mathbbm,prl]{revtex4}

\usepackage{amsmath,amsfonts,amssymb,graphics,graphicx,epsfig,color,times,bbm}

\begin{document}
\bibliographystyle{apsrev}

\title{Comment on ``Separability of quantum states and the violation of Bell-type inequalities''}

\author{Christoph Simon}
\email{christoph.simon@ujf-grenoble.fr}
 \affiliation{
 Laboratoire de Spectrom\'{e}trie
Physique, CNRS et Universit\'{e} de Grenoble 1, St.\ Martin
d'H\`{e}res, France}

\date{\today}

\begin{abstract}
The statement of E.R. Loubenets, Phys. Rev. A {\bf 69}, 042102
(2004), that separable states can violate classical probabilistic
constraints is based on a misleading definition of classicality,
which is much narrower than Bell's concept of local hidden
variables. In a Bell type setting the notion of classicality used
by Loubenets corresponds to the assumption of perfect correlations
if the same observable is measured on both sides. While it is
obvious that most separable states do not satisfy this assumption,
this does not constitute ``non-classical'' behaviour in any usual
sense of the word.
\end{abstract}

\pacs{}

\maketitle

In Ref. \cite{loubenets} the author claims to show that separable
quantum states do not satisfy all ``classical probabilistic
constraints''. Here I argue that the definition of classicality
used is misleading.

The notion of ``joint classical measurements'', on which the
author's assertion rests, is explicitly introduced in the appendix
of Ref. \cite{loubenets}. She considers three ``properties'' $A$,
$D_1$ and $D_2$, either two of which can be measured jointly. In a
Bell-type situation with two separate subsystems (to which the
formalism is applied in the paper), these properties correspond to
measurements on a single subsystem. The author considers the joint
measurements $A-D_1$, $A-D_2$ and $D_1-D_2$. She does not specify
which of the two properties is measured for which subsystem. In
fact, it is implicit in her definition that this does not matter.

This becomes clear from the author's postulates for the
correlation functions corresponding to the joint measurements, see
Eq. (A3) in the appendix of Ref. \cite{loubenets}. She assumes
that there are three functions $f_A(\theta)$, $f_{D_1}(\theta)$
and $f_{D_2}(\theta)$ of the hidden variables $\theta$, which give
the measured values of the properties $A$, $D_1$ and $D_2$. It is
obvious from Eq. (A3) that the same function is supposed to apply,
independently of the subsystem on which the respective property is
measured. In order for the three joint measurements to be
possible, at least one of the three properties has to be measured
on the first subsystem in one combination and on the second
subsystem in another combination, i.e. it has to switch sides. Eq.
(A3) implicitly assumes that this makes no difference.

In a general local hidden variable model as considered by Bell in
Ref. \cite{bell} a priori there are six independent functions,
corresponding to measuring each property $A$, $D_1$ and $D_2$
either on the first or on the second subsystem. Only if one has
additional information about the system under consideration can
this number be reduced. For example, in Ref. \cite{bell} Bell
assumed that the system under consideration exhibits perfect
anti-correlations, if the same property is measured on both
subsystems. This was made possible by the fact that he only wanted
to apply his inequality to the quantum mechanical singlet state of
two spin-(1/2) systems, which exhibits such perfect
anti-correlations.

Eq. (A3) in Ref. \cite{loubenets} corresponds to the assumption of
perfect correlations. For a given value of $\theta$, i.e. for a
given system, the same value will be obtained for the property
$A$, irrespectively of whether it is measured on the first or on
the second subsystem. If it is measured for both subsystems at the
same time, there will thus be perfect correlation.

In contrast to Bell, the author then applies the derived
inequality to a state which does not exhibit such perfect
correlations at all. In fact, if the same observable (the spin
along the same direction) is measured on both sub-systems for the
state Eq. (6) of Ref. \cite{loubenets}, the spin correlation
function for the given state is always {\it non-positive}, as can
be seen from Eq. (7) of Ref. \cite{loubenets}, whereas perfect
correlations would correspond to a value $+1$ of the correlation
function for all such joint measurements along identical
directions. It is thus not at all surprising that the state
violates the derived inequality.

However, it is quite misleading to suggest that this constitutes a
violation of ``classical probabilistic constraints''. It is simply
a consequence of the fact that the author has implicitly defined
``classicality'' to mean the existence of perfect correlations.
There are of course many examples of classical systems in the
conventional sense that do not exhibit perfect correlations.

In conclusion, Ref. \cite{loubenets} does not demonstrate
non-classical properties of separable states in any usual sense of
the word. It is worth mentioning that in Ref. \cite{bell} Bell
gives an explicit hidden variable model for spin measurements on a
single spin-(1/2) system, which immediately leads to a local
hidden variable model for separable states of two spin-(1/2)
systems such as the state discussed by the author.

\end{document}